\documentclass[%
 reprint,
 amsmath,amssymb,
 prl,
]{revtex4-1}
\usepackage{wasysym}
\usepackage{graphicx}
\usepackage{dcolumn}
\usepackage{bm}
\DeclareMathAlphabet{\mathpzc}{OT1}{pzc}{m}{it}

\begin{document}

\preprint{APS/123-QED}

\title{Correlation Effects in Quadrupole Insulators: a Quantum Monte Carlo Study}

\author{Chen Peng}
\affiliation{Department of Physics, Renmin University of China, Beijing 100872, China}

\author{Rong-Qiang He}\email{rqhe@ruc.edu.cn}
\affiliation{Department of Physics, Renmin University of China, Beijing 100872, China}

\author{Zhong-Yi Lu}\email{zlu@ruc.edu.cn}
\affiliation{Department of Physics, Renmin University of China, Beijing 100872, China}

\date{\today}

\begin{abstract}
The quadrupole insulator, a high-order topological insulator, with on-site Hubbard interaction is numerically studied by large-scale projector quantum Monte Carlo (PQMC) simulations. The Green's function formalism is successfully used to characterize topological properties in interacting quadrupole insulators for the first time. We find that the topological quadrupole 
insulator is stable against weak interactions and turns into a trivial antiferromagnetic (AFM) insulator by a continuous topological phase transition (TPT) for strong interactions. The critical exponents related to the TPT are estimated to be $\nu=0.67(4)$, $\beta=0.40(2)$, which are distinct from those of the known AFM transitions and suggest a new universality class.
\end{abstract}

\maketitle


The bulk-edge correspondence \cite{PhysRevLett.89.077002} is one of the most fundamental concepts for understanding topological insulators (TIs) in $d$-dimensional systems which possess gapless modes in their ($d-1$)-dimensional boundaries \cite{PhysRevLett.98.106803,PhysRevLett.95.146802,PhysRevB.76.045302}. Recently, topological crystalline insulators generalize this bulk-edge correspondence to high-order topological insulators (HOTIs) \cite{PhysRevB.96.245115,Benalcazar61,PhysRevLett.119.246402,PhysRevLett.122.086804}. In addition, the concept of HOTIs is also generalized to high-order topological superconductors \cite{HOTI.SC.s0.PhysRevB.99.125149,HOTI.SC.s1.PhysRevLett.121.186801,HOTI.SC.s2.PhysRevLett.121.096803,HOTI.SC.s3.PhysRevLett.119.246401,PhysRevLett.119.246401}. The predicted Majorana corner/hinge states in such systems are promising building blocks (qubits) for realizing quantum computation \cite{computation1.bomantara:2019,computation2.pahomi:2019}.

Attracted by the novel properties and potential applications of HOTIs, many theoretical works have been done to predict HOTIs in realistic systems, such as quasicrystals, sonic crystals, non-hermitian systems, and solid state systems \cite{HOTI.predict1,HOTI.predict2,HOTI.predict3,HOTI.predict4,HOTI.predict5,HOTI.predict6,HOTI.predict7,HOTI.predict8}. And some experimental works have even shown the direct observation of HOTIs in photonic/phononic crystals, electrical circuits as well as crystalline solids, such as bismuth \cite{HOTI.exp1,HOTI.exp2,HOTI.exp3,HOTI.exp4,HOTI.exp5,HOTI.exp6}.

Unlike bosonic systems, fermionic systems with spin degrees of freedom are subjected to the Coulumb repulsion which may give rise to many exotic phenomena \cite{PhysRevB.88.085406,PhysRevB.91.115317,PhysRevLett.120.157205,PhysRevB.26.4278,PhysRevB.84.195107,PhysRevB.93.115150,PhysRevB.97.081110,PhysRevLett.106.100403}. However, there are only a few studies which discuss correlation effects  in HOTIs \cite{HOTI.interaction1,HOTI.interaction2,HOTI.interaction3}. Unbiased numerical research on interacting HOTIs is almost a blank and is the main motivation of this work.

In this paper, we focus on such a family of HOTIs, which are called quadrupole topological insulators. The model is constructed in a non-interacting spinless system with only nearest neighbor hopping terms in two dimensions \cite{Benalcazar61,PhysRevB.96.245115}. In conventional $d$-dimensional TIs, $(d-1)$-dimensional edge modes can be understood by electric dipoles and charge pumps, which are related to the Wilson loops and energy bands. However, the bulk topology of multipole insulators need to be understood by electric multipole moments and dipole pumps, which are respectively related to the nested Wilson loops and Wannier bands, respectively. The two-dimensional quadrupole insulator studied here has a vanishing dipole moment and a quantized quadrupole moment in its bulk. Then, the bulk quadrupole moment in a finite sample gives rise to dipole moments on its one-dimensional edges and zero-dimensional corner charges. In order to discuss correlation effects in quadrupole insulators, we introduce spin degrees of freedom and on-site Hubbard interactions to quadrupole insulators and simulate the system using sign-problem-free projector quantum Monte Carlo (PQMC) method. 

\begin{figure}[b]
  \includegraphics[scale=0.75]{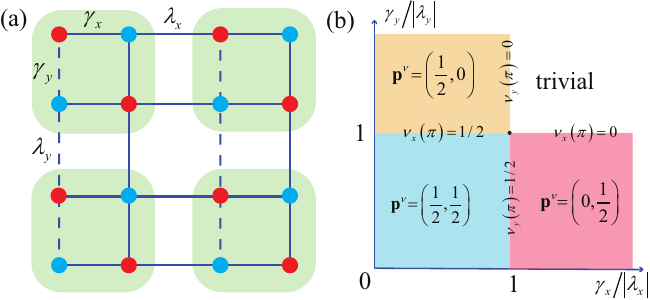}
  \caption{(a) Schematic of the lattice. Dashed lines represent the hopping terms with a negative sign. (b) Phase diagram of the model described by Eq.(\ref{subeq:2}) with $U=0\lambda$ (we set $\lambda_x=\lambda_y=\lambda$ as energy unit in our simulations). Only the first quadrant of the parameter space is shown, compared with Ref.\cite{Benalcazar61}. Quadrupole phase survives in the blue area. Other two topological phases marked by yellow and red possess non-trivial dipole moments in $x$ and $y$ edges, respectively. Wannier gaps close at the phase boundaries between different topological phases. Dirac semi-metal with parameters $(\gamma_x,\gamma_y)/{ \lambda} = (1, 1)$ is marked by black dot, which can directly connected the quadrupole phase and a trivial band insulator.}
  \label{fig0}
\end{figure}

\begin{figure}[b]
  \includegraphics[scale=0.13]{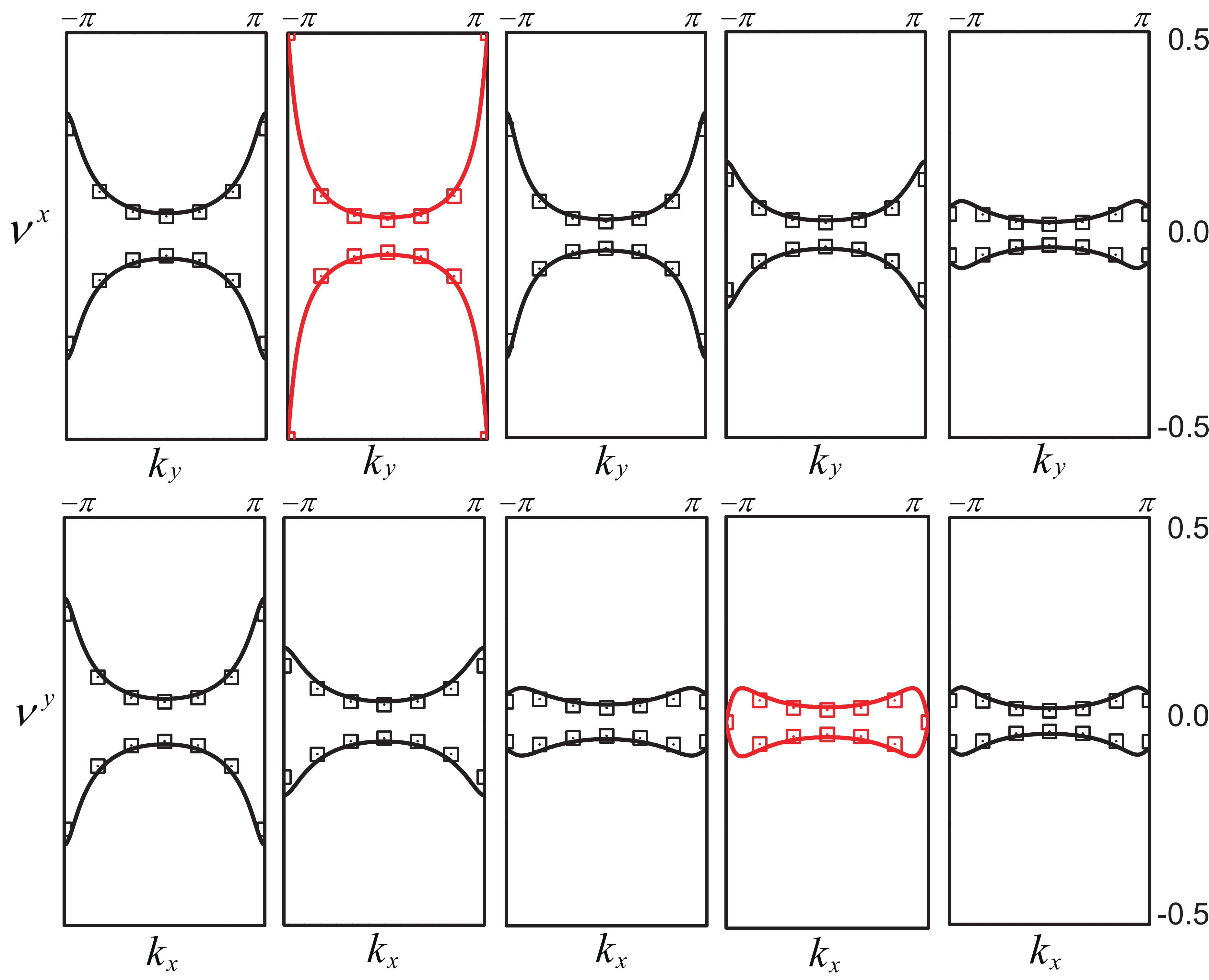}
  \caption{Wannier bands $\nu_x^{\pm}(k_y)$ and $\nu_y^{\pm}(k_x)$ in domain $[-\pi,\pi]$. The five columes from left to right correpspond to the path in parameter space $(\gamma_x,\gamma_y)/\lambda$ : $(0.75,0.75) \to (0.75,1) \to (0.75,1.25) \to (1,1.25) \to (1.25,1.25)$, respectively. Wannier bands $\nu_x$ are shown in upper panel while $\nu_y$ are shown in the lower panel. The continuous lines are calculated by $H_0({\bf{k}})$ with lattice size $L=100$ for the non-interacting system, while the square symbols are calculated by PQMC using the Green's function formalism with lattice size $L=6$ and $U=2\lambda$.}.
  \label{fig1}
\end{figure}
{\em{ Model and method.}}---Together with a concrete spinless lattice model with just nearest neighbor hopping terms defined in a square lattice, the concept of electric multipole moments in solid state systems is first established in Ref.\cite{Benalcazar61}. In order to study the correlation effects in quadrupole insulators, we stack up two copies of this model and introduce spin degrees of freedom ($\uparrow$ and $\downarrow$) to each copy, respectively. Mirror symmetries $M_x$ and $M_y$, which protect topological phases, still survive, since we do not introduce spin-orbital coupling terms. So, it is enough to characterize this spinful system by simply discussing topological properties carried by only one spin flavor. After adding the on-site Hubbard interaction, which accounts simply for the Coulumb repulsion, the model studied here is described by $H = H_{0} + H_I$:
\begin{eqnarray}
  {H_0} =&& \sum\limits_{{\bf{R}}\sigma } {[ {{\gamma _x}\left( {c_{{\bf{R}},1,\sigma }^ + c_{{\bf{R}},3,\sigma }^{} + c_{{\bf{R}},2,\sigma }^ + c_{{\bf{R}},4,\sigma }^{} + {\rm{H}}{\rm{.c}}{\rm{.}}} \right)}} 
  \nonumber
  \\
  && { + {\gamma _y}\left( {c_{{\bf{R}},1,\sigma }^ + c_{{\bf{R}},4,\sigma }^{} - c_{{\bf{R}},2,\sigma }^ + c_{{\bf{R}},3,\sigma }^{} + {\rm{H}}{\rm{.c}}{\rm{.}}} \right)}
  \nonumber
  \\
  && { + {\lambda _x}\left( {c_{{\bf{R}},1,\sigma }^ + c_{{\bf{R}} + \hat x,3,\sigma }^{} + c_{{\bf{R}},4,\sigma }^ + c_{{\bf{R}} + \hat x,2,\sigma }^{} + {\rm{H}}{\rm{.c}}{\rm{.}}} \right)}
  \nonumber
  \\
  && { + {\lambda _y}\left( {c_{{\bf{R}},1,\sigma }^ + c_{{\bf{R}} + \hat y,4,\sigma }^{} - c_{{\bf{R}},3,\sigma }^ + c_{{\bf{R}} + \hat y,2,\sigma }^{} + {\rm{H}}{\rm{.c}}{\rm{.}}} \right)}],
  \nonumber
  \\
  {H_I} =&& \frac{U}{2}\sum\limits_i {\left( {{n_{i \uparrow }} + {n_{i \downarrow }} - 1} \right)}^2.
  \label{subeq:2}
\end{eqnarray}
Wannier-sector polarizations $p_{x}^{\nu_y^{\pm}}(p_{y}^{\nu_x^{\pm}})$, which describe the electronic polarizations along $x(y)$ within Wannier bands $\nu_y^{\pm}(\nu_x^{\pm})$, are quantized to $1/2$ or $0$ due to the mirror symmetries $M_x$ and $M_y$, where $\nu_x^{\pm}$($\nu_y^{\pm}$) stands for the upper ($+$) or lower ($-$) Wannier band along $k_x$($k_y$). (More details in Supplemental Material.) Then, different topological phases can be distinguished by the topological index ${\bf{P}}^{\nu}=(p_x^{\nu_y^-},p_y^{\nu_x^-})$ (Fig.\ref{fig0}(b)). Wannier values and the locations of Wannier gap closing point at each phase boundary are also given in Fig.\ref{fig0}(b).

\begin{figure}[b]
  \includegraphics[scale=0.3]{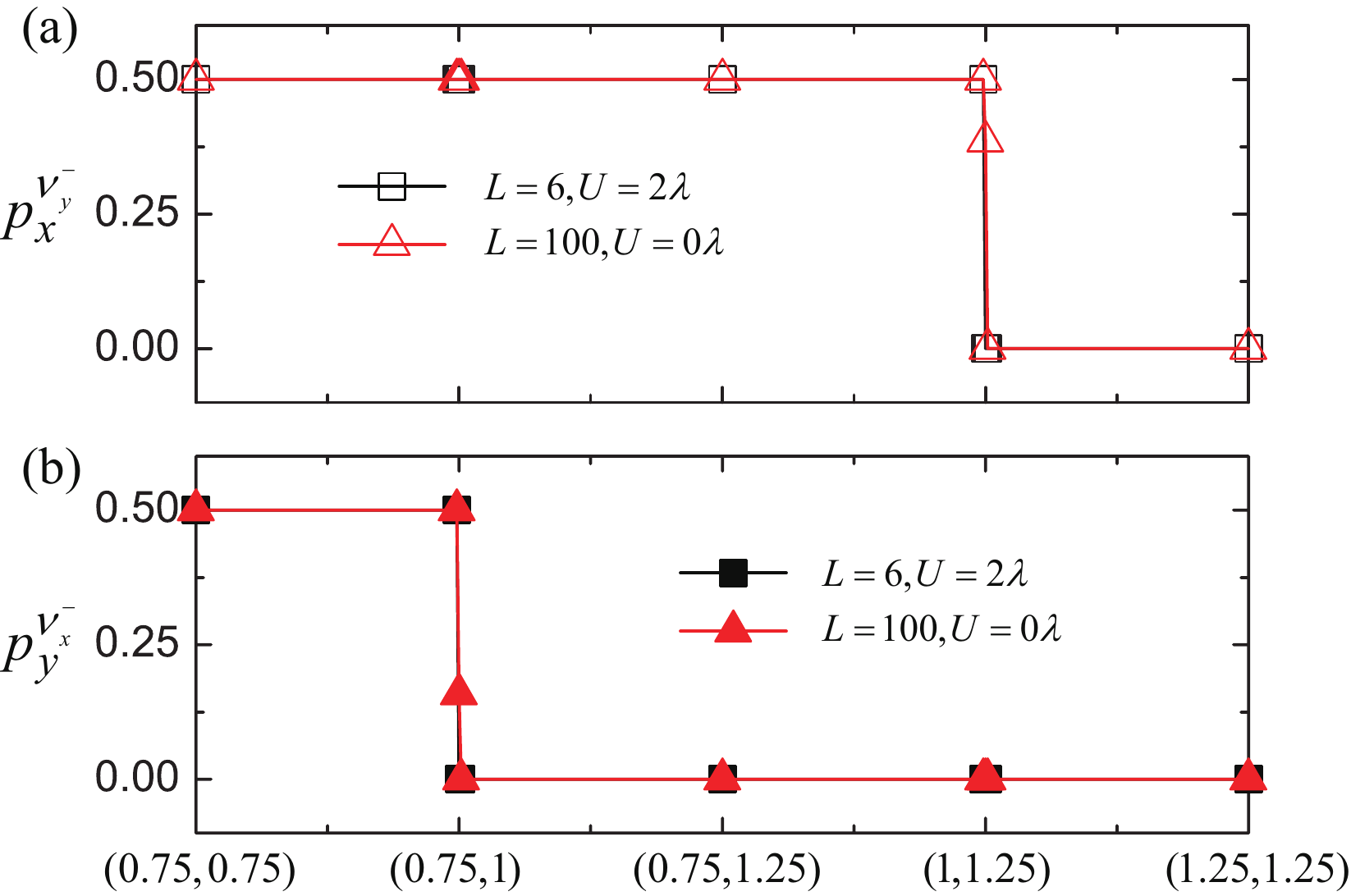}
  \caption{Topological index ${\bf{P}}^{\nu}=(p_x^{\nu_y^-},p_y^{\nu_x^-})$ calculated through the same path in parameter space as in Fig.\ref{fig1}. Black and red symbols stand for interacting case ($U=2\lambda,L=6$) and non-interacting case ($U=0\lambda,L=100$), respectively.}
  \label{fig2}
\end{figure}

Setting parameters $\lambda_x = \lambda_y = \lambda$ as energy unit, we investigate the model introduced above at half filling using large-scale PQMC simulation, which is sign-problem free as long as $H_0$ includes only nearest-neighbour hopping terms \cite{Assaad2008}. Another issue we need to address is to distinguish different topological phases in a correlated system. In the conventional TIs, it is the Green's function formalism that plays a key role in identifying different topological phases after introducing interactions \cite{PhysRevX.2.031008,Wang_2013,PhysRevLett.105.256803,PhysRevX.2031008,PhysRevX.9.011012}. In this paper, We simply use the zero-frequency Green's function $G^{-1}({\bf{k}},i\omega=0)$ to replace the non-interacting bloch Hamiltonian $H_0(\bf{k})$ in the calculation of the topological index $\bf{P}^\nu$. (cf. Supplemental Material.) 


{\em{Single-particle level TPTs with interactions.}}---In order to investigate correlation effects in quadrupole insulators, we need to verify whether the zero-frequency Green's function mentioned above can characterize the topological properties in quadrupole insulators. Unlike in the conventional topological insulators, topological properties in quadrupole insulators studied here are encoded in the Wannier bands. In other words, the topological phase is protected by the mirror symmetries and can not transform to another phase as long as the Wannier gaps, rather than the energy bands for the conventional topological insulators, keep open. So, we first investigate whether TPTs can still be detected by the Wannier gap closing events after introducing on-site Hubbard interaction as in the non-interacting case.

The TPTs at the single-particle level can be driven by a pair of hopping parameters $(\gamma_x,\gamma_y)/\lambda$ as shown in Fig.\ref{fig0}(b). We choose the path in parameter space, $(0.75,0.75) \to (0.75,1) \to (0.75,1.25) \to (1,1.25) \to (1.25,1.25)$, which is discussed in Ref.\cite{PhysRevB.96.245115}. Starting from the quadrupole phase characterized by ${\bf{P}^{\nu}}=(1/2,1/2)$, we increase $\gamma_y/\lambda$ from 0.75 to 1.25 while keeping $\gamma_x / \lambda$ fixed at 0.75. As the quadrupole phase approaches the TPT, the Wannier bands $\nu_x(k_y)$ become gapless at $k_y=\pi$ as they become twofold degenerate at $\nu_x(\pi) = 1/2$. After this TPT, the Wannier gap reopens, and the ground state is still in a non-trivial topological phase (with ${\bf{P}^\nu} = (1/2, 0)$) but not a quadrupole phase. Then with keeping $\gamma_y$ fixed at $\gamma_y / \lambda = 1.25$ and increasing $\gamma_x$, the gap of another Wannier band $\nu_y(k_x)$ is closing at $k_x=\pi$ at the transition point $(1,1.25)$. Finally, the electric polarization $p_x^{\nu_y^-}$ within Wannier band $\nu_y^-$ changes from $1/2$ to $0$, which means the ground state is in a completely trivial phase with ${\bf{P}^{\nu}}=(0,0)$.

\begin{figure}[b]
  \includegraphics[scale=0.32]{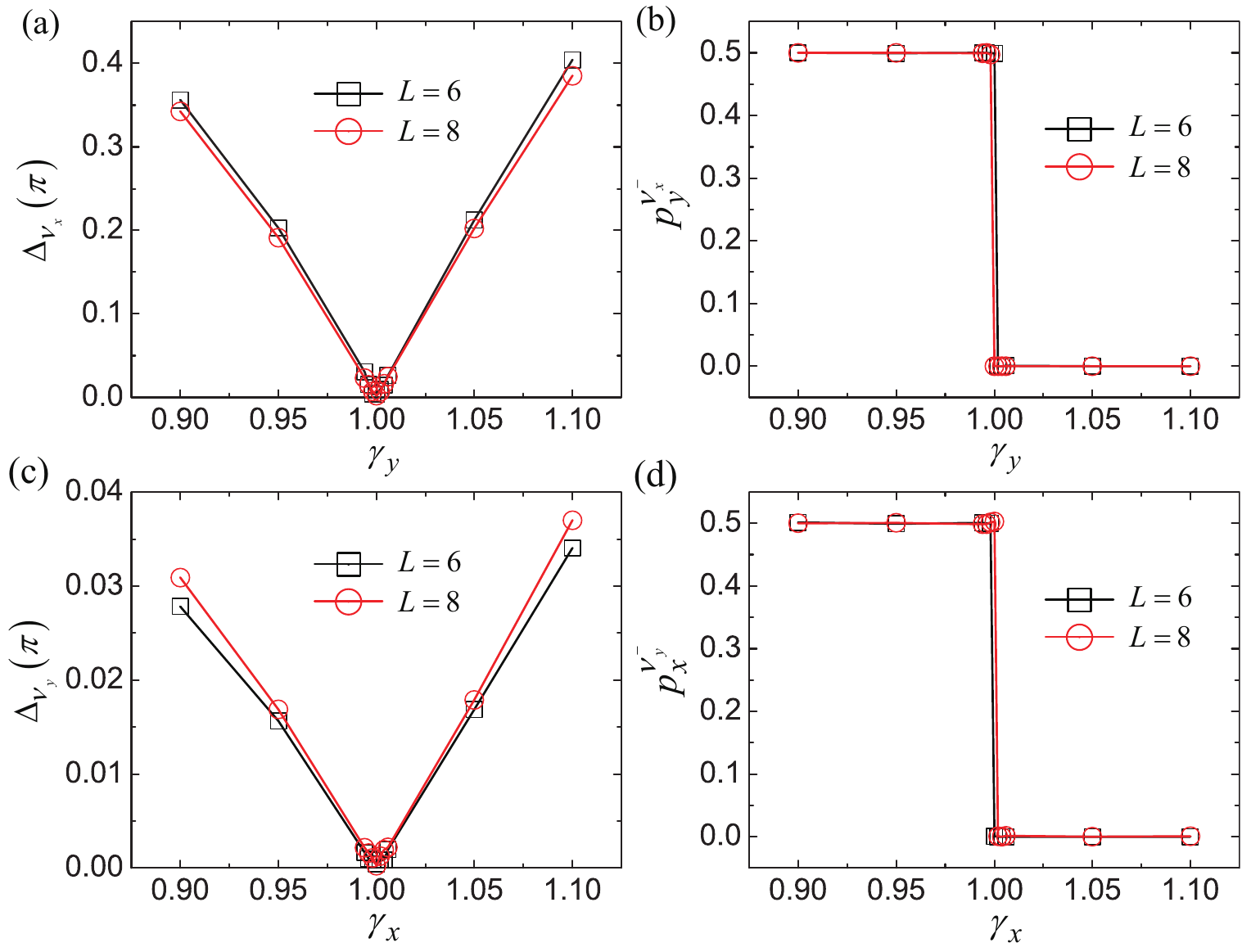} 
  \caption{Wannier-band gaps and topological index $\bf{P}^{\nu}$ around the phase boundaries with $U=8\lambda$. (a) Wannier-band gaps $\Delta_{\nu_x}(k_y=\pi)$ and (b) Wannier-sector polarizations $p_y^{\nu_x^-}$,$p_x^{\nu_y^-}$ are calculated by setting $\gamma_x= 0.8\lambda$ and increasing $\gamma_y$ from $0.9\lambda$ to $1.1\lambda$. While (c) $\Delta_{\nu_y}(k_x=\pi)$ and (d) $p_x^{\nu_y^-}$ are calculated by choosing $\gamma_y=1.2\lambda$ and sweeping $\gamma_x$ from $0.9\lambda$ to $1.1\lambda$.}
  \label{fig3}
\end{figure}

The evolution of Wannier bands through the path mentioned above are calculated by the PQMC using the Green's function formalism. Numerical results are shown by square symbols in Fig.\ref{fig1} with system size $L=6$ and interaction strength $U=2\lambda$. We find that the Wannier bands $\nu_x$ and $\nu_y$ experience a gap closing and reopening process through the TPT similar to the non-interacting case. To make a concrete contrast, we also calculate Wannier bands and Wannier-sector polarizations with $U=0$ in the same path. Numerical results are shown in Fig.\ref{fig1} by pure lines. Since system the interacting case with $U=2\lambda$ is not far from the non-interacting case, we can find that the evolution of Wannier bands in the interacting system through the path in parameter space only slightly deviate from that of the non-interacting case. What's more, we find that the Wannier gaps calculated by $G({\bf{k}},i\omega=0)$ with finite $U$ also experience a gap closing and reopening process through the phase boundaries. These results show that the TPTs happen at $(0.75,1.0)$ and $(1,1.25)$ and also mean that TPTs can be successfully detected in interacting quadrupole insulators by our method. 

To further confirm the validity of the Green's function formalism, we calculate topological index $\bf{P}$ by definition as a direct indication of TPTs (Fig.\ref{fig2}). Still in the system with $U=2\lambda$, we find Wannier-sector polarizations $p_x^{\nu^-}$ and $p_y^{\nu^+}$ jump at $(1,1.25)$ and $(0.75,1)$ respectively, which is consistent with the results we derived from the Wannier bands argument. Thus, the topological properties in interacting quadrupole insulators can be correctly captured by $G({\bf{k}},i\omega=0)$. 

\begin{figure}[b]
  \includegraphics[scale=0.33]{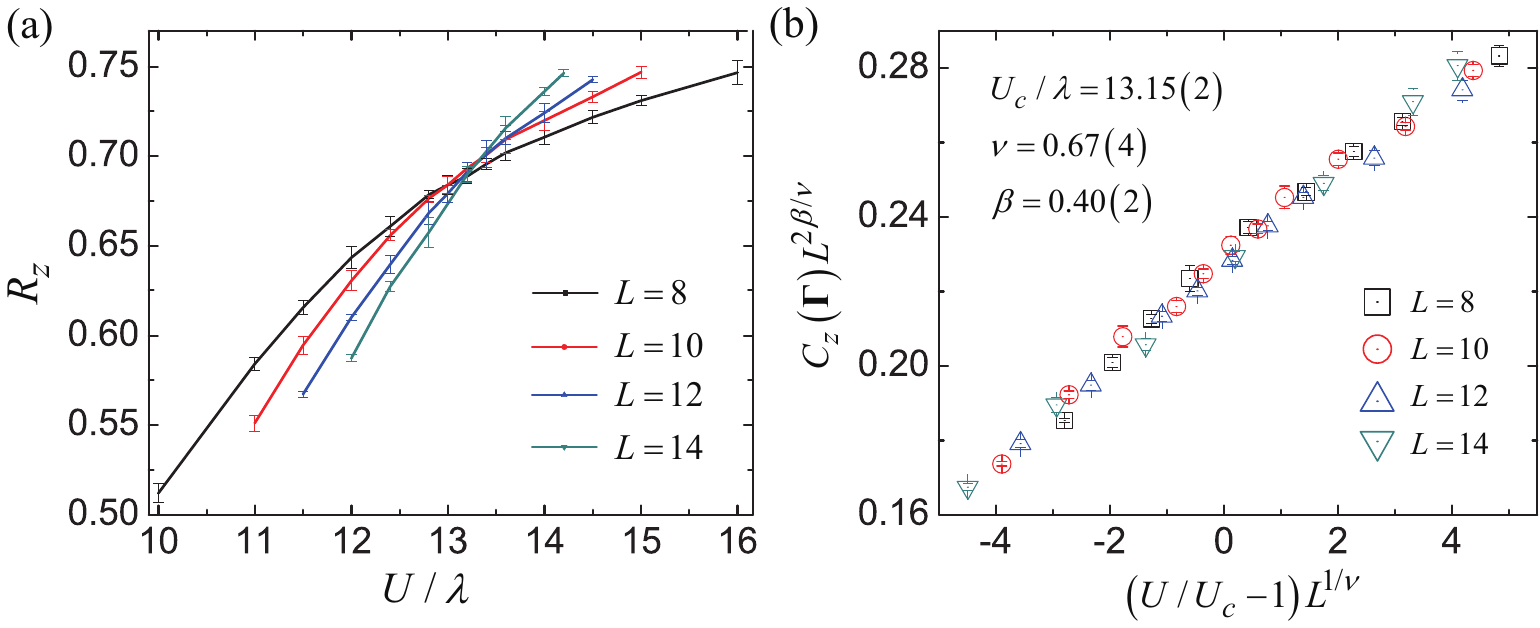} 
  \caption{(a) Correlation ratios $R_z(U,L)$ with different sizes and interaction strengths. The crossing point gives an estimate of the critical point $U_c$. (b) Data collapse of structure factors $C_z(U,L)$, which gives $U_c/\lambda=13.15(2)$, $\nu = 0.67(4)$ and $\beta = 0.40(2)$. The calculations is done with $(\gamma_x,\gamma_y)/\lambda=(0.8,0.8)$. }
  \label{fig4}
\end{figure}

In order to investigate how the phase boundaries vary with interactions, we increase the interaction strength up to $U=8\lambda$. We first investigate the transition between ${\bf{P}}^{\nu}=(1/2,1/2)$ and ${\bf{P}}^{\nu}=(1/2,0)$ by fixing $\gamma_x=0.8\lambda$ and increasing $\gamma_y$ from $0.9\lambda$ to $1.1\lambda$. Wannier bands $\nu_x$ close the gap at $k_y=\pi$ and Wannier-sector polarization $p_y^{\nu_x^-}$ jumps from $1/2$ to $0$ at the same point $\gamma_y=\lambda$ as in the non-interacting case (Figs.\ref{fig3}(a,b)). Similar results are also shown in the phase transition from ${\bf{P}}^{\nu}=(1/2,0)$ to ${\bf{P}}^{\nu}=(0,0)$ with increasing $\gamma_x$ and fixing $\gamma_y=1.2\lambda$ (Figs.\ref{fig3}(c,d)). Unlike in the conventional topological insulators \cite{PhysRevB.87.121113,PhysRevB.89.235104}, the phase transitions between different topological phases of Wannier bands are not affected by the Hubbard interaction. The stability of phase boundaries may be accounted for by the novel definition of quadrupole insulators that the topological properties are encoded in Wannier bands rather than energy bands.

{\em{Interaction induced TPTs.}}---Although we have studied correlation effects in quadrupole insulators, TPTs discussed above still stay at the single-particle level as the phase transition is driven just by tight-binding parameters. A direct route to construct an interaction driven TPT in our model is increasing the on-site Hubbard $U$ until an AFM long-range order is formed. Here, we study the transition from the quadrupole phase to AFM insulator (AFMI) and find that they are continuous. The Wannier-sector polarization $p_x^{\nu^-}$ is calculated and our results clearly reveal that the bulk quadrupole moment is gradually destroyed by the AFM order.

\begin{figure}[b]
  \includegraphics[scale=0.33]{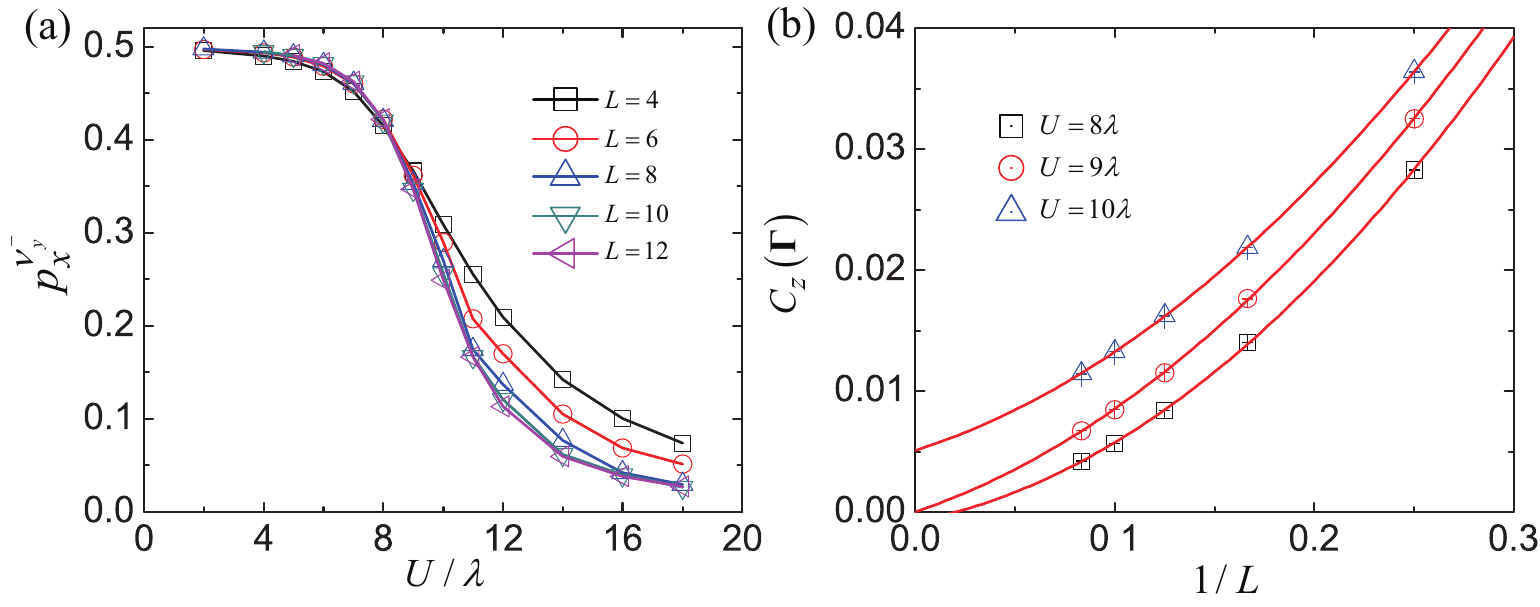} 
  \caption{(a) Topological index $p_x^{\nu_y^-}$ with different lattice sizes. (b) Finite-size extrapolation of $z$-direction magnetic structure factors of $L=4,6,8,10,12$ systems by quadratic polynomials. $(\gamma_x/\lambda,\gamma_y/\lambda)=(0.8,0.8)$. A very small staggered magnetic field $\mu = 0.001\lambda$ is introduced to the system.}
  \label{fig5}
\end{figure}

In order to characterize the quadrupole insulator to AFMI transition, we calculate the correlation ratio of AFM order parameter: ${R_z}\left( {U,L} \right) = 1 - \frac{{{C_z}\left( {{\bf{\Gamma}} + \delta {\bf{q}}} \right)}}{{{C_z}\left( {\bf{\Gamma}} \right)}}$ where ${C_z}\left( {\bf{\Gamma}} \right)$ is the structure factor ${C_z}\left( {\bf{\Gamma}}  \right) = \frac{1}{{4N^2}}\sum\nolimits_{ij\gamma } {\left\langle {S_{i\gamma }^zS_{j\gamma }^z} \right\rangle }$ where $i,j=1,2,...,N$ run over all unit cells and $\gamma$ stands for sub-lattice index in a unit cell. We choose parameters $(\gamma_x,\gamma_y)/\lambda=(0.8,0.8)$, which guarantee the ground state is a quadrupole insulator. Correlation ratios $R_z(U,L)$ with different system sizes cross at nearly the same point, which implies the transition is continuous since $R_z(U,L)$ is a renormalization invariant quantity in a continuous phase transition (Fig.\ref{fig4}(a)). Further, we make a data collapse to find the critical exponents by assuming $C_z({\bf{\Gamma}},U,L)L^{2\beta/\nu}=f_C((U/U_c-1)L^{1/\nu})$. The data collapse process is to find $\nu$ and $\beta$ to make all data points collapse at one single unknown curve described by function $f_C$. Results with optimized exponents are shown in Fig.\ref{fig4}(b). Critical exponents $\nu=0.67(4)$, $\beta=0.40(2)$ and critical point $U_c=13.15(2)\lambda$ is obtained, which reveals the TPT is actually continuous (Details to determine those values are given in Supplemental  Material). These exponents should be compared with those of the 3D-Heisenberg universality class, the 2D dimerized Heisenberg model, and the deconfined quantum criticality in VBS-AFM transitions (found in $J_1$-$J_2$ and $J$-$Q$ models) \cite{O3class,dimerizedheisenbergmodel,jq,j1j2}. The critical exponents obtained here are inconsistent with those in these three cases, and suggest a new universality class for TPTs between quadrupole insulators and and AFM insulators.

Having known that the AFM phase is formed around $U_c \sim 13.15\lambda$, the interplay between the bulk topology and AFM order is still uncertain. The physics underlying the TPT here can be understood by spontaneous symmetry breaking. Different topological phases described by index ${\bf{P}}^{\nu}$ are well-defined only when preserving mirror symmetries $M_x,M_y$ \cite{Benalcazar61}. In other words, Wannier-sector polarizations $p_x^{\nu_y^{\pm}},p_y^{\nu_x^{\pm}}$ are no longer quantized to $1/2$ or $0$ when mirror symmetry is spontaneously broken by AFM order. Meanwhile, the Wannier gap closing mechanism, which is used to describe TPT, also fails since the topological properties encoded in different Wannier sectors $\nu^+$ and $\nu^-$ can mess up without closing the Wannier band gaps. However, in previous studies, topological index failed to describe the destruction of topology by long-range order in numerical simulations, as spontaneous symmetry breaking did not happen in a finite size system or due to the breakdown of the Green's function formalism \cite{PhysRevB.87.205101,PhysRevB.93.195164}. 

Hence, in order to visualize the competition between topology and AFM order directly, we simulate the system with manually introducing a very small staggered magnetic field, which can break the mirror symmetries. Specifically, we add positive or negative magnetic fields with strength $\mu = 0.001\lambda$ to our system corresponding to the sites marked with red or blue in Fig.\ref{fig0}(a). Since the AFM order plays the same role on $p_x^{\nu_y}$, $p_y^{\nu_x}$ and the constraint due to the inversion symmetry $\mathcal{I}$ are still valid \cite{Benalcazar61} (details in Supplemental Material), we only use one of them to characterize the topology in the quadrupole phase in Fig.\ref{fig5}(a). Far from the discontinuous behavior in TPTs at the single-particle level, electric polarization $p_x^{\nu_y^-}$ shows a smooth decrease from 1/2 to 0, which means the topological order is suppressed by the AFM order gradually as we expect. By extrapolating structure factor $C_z(\Gamma)$ to the thermodynamic limit with different sizes $L=4,6,8,10,12$, we find that the AFM order appears around $U=9\lambda$, which is smaller than $U_c\sim13.15\lambda$, since spins are polarized toward $z$-direction under the small external staggered magnetic field $\mu$ (Fig.\ref{fig5}(b)).

From directly monitoring the topological index around the TPT with introducing a tiny staggered magnetic field, we can clearly find the competition between AFM and topological order. The suppression of non-trivial topology happens simultaneously with the formation of AFM order, which means the AFM phase obtained here is a simple AFMI. 

{\em{\label{section5}Summary.}}---We apply the Green's function formalism to study interacting topological quadrupole insulators numerically for the first time. Through our numerical results using PQMC, we find that this formalism can successfully capture all the topological properties encoded in Wannier bands even when interactions are introduced. For the single-particle level TPTs, due to the novel machenism of TPTs in quadrupole insulators, the phase boundaries between different topological phases marked by ${\bf{P}}^{\nu}$ keep unchanged with increasing the strength of on-site Hubbard interaction up to $U=8\lambda$.

Besides the single-particle level TPTs, which have non-interacting counterpart, we also study a truly interaction driven TPT quantitatively. A continuous phase transition is found between the quadrupole insulator and AFMI. Critical exponents $\nu,\beta$ are obtained by data collapse, which suggest a new universality class for HOTIs. By adding a tiny staggered magnetic field, the competition between topology and AFM order is revealed through the smooth decreasing of topological index $p_x^{\nu_y^-}$ from 1/2 to 0. This indicates that the topological order is destroyed by AFM simultaneously and the AFM phase is a trivial Mott insulator.

This work was supported
by the National Natural Science Foundation of China (Grants No. 11774422 and No. 11874421). R. Q. H. was supported by the Fundamental Research Funds for the Central Universities, and the Research Funds of Renmin University of China (Grant No. 18XNLG11). Computational resources were provided by
the Physical Laboratory of High Performance Computing at Renmin University of China and National Supercomputer Center in Guangzhou with Tianhe-2 Supercomputer.

\nocite{*}

\bibliography{myref}

\newpage

\setcounter{equation}{0}
\renewcommand{\theequation}{S\arabic{equation}}
\setcounter{figure}{0}
\renewcommand{\thefigure}{S\arabic{figure}}
\setcounter{section}{0}
\renewcommand{\thesection}{S\arabic{section}}
\setcounter{table}{0}
\renewcommand{\thetable}{S\arabic{table}}
\onecolumngrid

\vskip1cm

\centerline{\large {\bf {Supplemental Material for }}}

\vskip1mm

\centerline{\large {\bf { ``Correlation Effects in Quadrupole Insulators: a Quantum Monte Carlo Study''}}}

\vskip8mm

\twocolumngrid
\section{\label{S0} solving the model with projector quantum Monte Carlo method}
Our model writes as $H=H_0 + H_I$:
\begin{eqnarray}
  {H_0} =&& \sum\limits_{{\bf{R}}\sigma } {[ {{\gamma _x}\left( {c_{{\bf{R}},1,\sigma }^ + c_{{\bf{R}},3,\sigma }^{} + c_{{\bf{R}},2,\sigma }^ + c_{{\bf{R}},4,\sigma }^{} + {\rm{H}}{\rm{.c}}{\rm{.}}} \right)}} 
  \nonumber
  \\
  && { + {\gamma _y}\left( {c_{{\bf{R}},1,\sigma }^ + c_{{\bf{R}},4,\sigma }^{} - c_{{\bf{R}},2,\sigma }^ + c_{{\bf{R}},3,\sigma }^{} + {\rm{H}}{\rm{.c}}{\rm{.}}} \right)}
  \nonumber
  \\
  && { + {\lambda _x}\left( {c_{{\bf{R}},1,\sigma }^ + c_{{\bf{R}} + \hat x,3,\sigma }^{} + c_{{\bf{R}},4,\sigma }^ + c_{{\bf{R}} + \hat x,2,\sigma }^{} + {\rm{H}}{\rm{.c}}{\rm{.}}} \right)}
  \nonumber
  \\
  && { + {\lambda _y}\left( {c_{{\bf{R}},1,\sigma }^ + c_{{\bf{R}} + \hat y,4,\sigma }^{} - c_{{\bf{R}},3,\sigma }^ + c_{{\bf{R}} + \hat y,2,\sigma }^{} + {\rm{H}}{\rm{.c}}{\rm{.}}} \right)}],
  \nonumber
  \\
  {H_I} =&& \frac{U}{2}\sum\limits_i {\left( {{n_{i \uparrow }} + {n_{i \downarrow }} - 1} \right)}^2.
  \label{eqS0:mdl}
\end{eqnarray}
Using the projection operator we can obtain the wave function of ground state $\left| {{\Psi _0}} \right\rangle  = {e^{ - \frac{\Theta }{2}H}}\left| {{\Psi _T}} \right\rangle$ and we choose as the trial wave function $\left| {{\Psi _T}} \right\rangle$ the Slater determinant describing the many-body ground state of $H_{0}$. The physical observables are measured by:
\begin{equation}
  \left\langle {\hat O} \right\rangle  = \frac{{\langle {\Psi _T}|{e^{ - \frac{\Theta }{2}H}}\hat O{e^{ - \frac{\Theta }{2}H}}\left| {{\Psi _T}} \right\rangle }}{{\langle {\Psi _T}|{e^{ - \Theta H}}\left| {{\Psi _T}} \right\rangle }}.
\end{equation}
To treat the interaction part, we divide the projection time into $M$ slices ($\Theta = M \Delta_\tau$) and make the Trotter decomposition to separate interaction term $H_I$ from $H$ in exponential:
\begin{equation}
  \langle {\Psi _T}|{e^{ - \Theta H}}\left| {{\Psi _T}} \right\rangle  = \langle {\Psi _T}|{\left( {{e^{ - \Delta _\tau ^{}{H_0}}}{e^{ - \Delta _\tau ^{}{H_I}}}} \right)^M}\left| {{\Psi _T}} \right\rangle  + O\left( {\Delta _\tau ^2} \right).
\end{equation}
 And after the Hubbard-Stratonovich (HS) transformation:
 \begin{equation}
{e^{ - \frac{{{\Delta _\tau }U}}{2}{{\left( {{n_{i \uparrow }} + {n_{i \downarrow }} - 1} \right)}^2}}} = \frac{1}{4}\sum\limits_{{s_i}} {\gamma \left( {{s_i}} \right){e^{\sqrt { - {\Delta _\tau }U} \eta \left( {{s_i}} \right)\left( {{n_{i \uparrow }} + {n_{i \downarrow }} - 1} \right)}}},
 \end{equation}
with $\gamma \left( { \pm 1} \right) = 1 + \sqrt 6 /3,\gamma \left( { \pm 2} \right) = 1 - \sqrt 6 /3,\eta \left( { \pm 1} \right) =  \pm \sqrt {2(3 - \sqrt 6 )} ,\eta \left( { \pm 2} \right) =  \pm \sqrt {2(3 + \sqrt 6 )} $, we can finally obtain the interaction terms in the formula with only fermion bilinears coupled to auxiliary fields. After performing a particle-hole transformation $c_{i \downarrow }^ +  \to {\left( { - 1} \right)^i}{c_{i \downarrow }}$, it is easy to check that the Monte Carlo weight is always semipositive. In the simulations, we set parameters $\lambda_x = \lambda_y = \lambda$ as energy unit and $\Delta_{\tau}=0.05/\lambda$. We choose $\Theta=35/\lambda,40/\lambda,45/\lambda,50/\lambda,60/\lambda$ to simulate system with size $L = 6,8,10,12,14$, respectively, which guarantees that the projected wave functions converge to ground states.

\section{\label{S1} detailed implementation to calculate $p_y^{\nu_x}$, $p_x^{\nu_y}$}

Since the model described in Eq.(\ref{subeq:2}) consists of four orbitals with spin degeneracy on square lattice, we take a 4-bands system as an example here to show the detailed implementation that we use to calculate Wannier-sector polarizations in the main text. 

A two-dimensional non-interacting system $H = \sum\limits_k {c_{k,\alpha }^ + {{\left[ {{h_k}} \right]}^{\alpha \beta }}{c_{k,\beta }}}$ can be diagonalized by a linear combination of operator $c_{k,\alpha}$:
\begin{equation}
  H = \sum\limits_{n,k} {\gamma _{n,k}^ + {\epsilon_{n,k}}{\gamma _{n,k}}} ,\quad \gamma _{n,k}^{} = \sum\limits_\alpha  {{{\left[ {u_k^{*n}} \right]}^\alpha }c_{k,\alpha }^{}},
  \label{eq:bloch-wave}
\end{equation}
where $n$ is band index. If we set periodic boundary condition with $N_x$ sites in $x$-direction, Wilson loop along $k_x$ can be defined within the subspace of occupied energy bands as:
\begin{equation}
{\mathcal{W}_x}\left( {\bf{k}} \right) = {G_{{\bf{k}} + \left( {N_x - 1} \right)\Delta {k_x}}}...{G_{{\bf{k}} + \Delta {k_x}}}{G_{\bf{k}}}.
\label{eq:wilson-loop}
\end{equation}
In the 4-band gapped system at half filling, $G_k$ is $2\times2$ matrix constructed by the eigenvectors of occupied bands obtained in Eq.(\ref{eq:bloch-wave}): ${\left[ {{G_{\bf{k}}}} \right]^{mn}} = \left\langle {{u_{{\bf{k}} + \Delta {k_x}}^m}}
\mathrel{\left | {\vphantom {{u_{{\bf{k}} + \Delta {k_x}}^m} {u_{\bf{k}}^n}}}
\right. \kern-\nulldelimiterspace}{{u_{\bf{k}}^n}} \right\rangle$, where $\left\langle {{u_{{\bf{k}} + \Delta {k_x}}^m}}
\mathrel{\left | {\vphantom {{u_{{\bf{k}} + \Delta {k_x}}^m} {u_{\bf{k}}^n}}}
\right. \kern-\nulldelimiterspace}{{u_{\bf{k}}^n}} \right\rangle = \sum\limits_\alpha  {{{\left[ {u_{{\bf{k}} + \Delta {k_x}}^{*m}} \right]}^\alpha }{{\left[ {u_{\bf{k}}^n} \right]}^\alpha }} $. In the thermodynamic limit, eigenvalues of each Wilson loop are just a phase $E_x^j(k_y) = e^{i2\pi \nu_x^j(k_y)}$. The phase factor $\nu_x^j(k_y)$ depends on the starting point of the Wilson loop is called Wannier center which interpreted as the center of electron distribution in the $x$ direction within a unit cell. After summing over Wannier centor with different occupied bands marked by $j$, we obtain dipole moments in each $k_y$: ${\bf{p}}_x(k_y)=\sum_j{\nu_x^j(k_y)}$. And then, the total dipole moment in the bulk of this 2D system is given by ${\bf{p}}_x=1/N_y\sum_{k_y}{\bf{p}}_x(k_y)$

We should notice that, the definition of the Wilson loop in Eq.(\ref{eq:wilson-loop}) also depends on the starting point of the loop $k_x$. Given a definite $k_y$, we can define $N_x$ Wilson loops $\mathcal{W}_x(n\Delta_{k_x},k_y)$ where $n=0,1,2...,N_x-1$. Although eigenvalues $\nu_x^j(k_y)$ are independent of the starting point $k_x$, Wannier functions $\left| {\Psi _R^j(k_y)} \right\rangle$ can be constructed by the eigenvectors of those Wilson loops with different starting points:
\begin{equation}
  \left| {\Psi _R^j(k_y)} \right\rangle  = \frac{1}{{\sqrt {N_x} }}\sum\limits_{n = 1}^{{N_{{\rm{occ}}}}} {\sum\limits_{k_x} {{{\left[ {\nu _{x,k_x}^j(k_y)} \right]}^n}{e^{ - ik_xR}}\gamma _{n,k_x,k_y}^ + } } \left| 0 \right\rangle,
  \label{eq:wannier-wave}
\end{equation}
where the $k_x$ dependent ${\left[ {\nu _{x,k_x}^j} \right]}^n$ is the $n$th component of the eigenvector of $2 \times 2$ Wilson loop $\mathcal{W}_x(k_x,k_y)$ with eigenvalue $\nu_x^j(k_y)$. 

In our 4-band example at half filling, we simply use $\pm$ to mark two values of index $j$. From a series of Wilson loops $\mathcal{W}_x(n\Delta_{k_x}, k_y)$, we can derive a one-to-one mapping between Wannier functions $\left| {\Psi _R^\pm(k_y)} \right\rangle$ and Wannier centers $\nu^\pm(k_y)$ for each $k_y$. Then we can make the definition of Wannier bands which describe the dispersion relation $\nu^\pm(k_y)$ versus $k_y$. After making the interchange of variables ${k_x} \leftrightarrow {k_y}$, we can derive another two Wannier bands which stand for $y$-direction polarizations $\nu^{\pm }_y(k_x)$. A vivid sketch of Wannier bands is shown in Ref.\cite{PhysRevB.96.245115}.

In the above statement, we first introduce the concept of Wilson loops defined in the subspace of occupied energy bands. Then, electronic polarizations of these bands can be derived from the eigenvalues of them. With the same procedure, we can further define the so-called nested Wilson loops based on each Wannier bands and the corresponding Wannier functions. For example, without providing tedious details, we directly give the expression of nested Wilson loop along $y$-direction within the subspace of Wannier band $\nu_x^+$ : 
\begin{eqnarray}
  \mathcal{W}_{y,{\bf{k}}}^{\nu _x^ + } = && \langle {w_{x,{\bf{k}} + {N_y}\Delta {k_y}}^ + } | {w_{x,{\bf{k}} + \left( {{N_y} \!-\! 1} \right)\Delta {k_y}}^ + }\rangle\langle {w_{x,{\bf{k}} + \left( {{N_y} \!-\! 1} \right)\Delta {k_y}}^ + } | ... 
  \nonumber 
  \\
  && |{w_{x,{\bf{k}} + \Delta {k_y}}^ + } \rangle \langle {w_{x,{\bf{k}} + \Delta {k_y}}^ + } | {w_{x,{\bf{k}}}^ + } \rangle,
\end{eqnarray}
where the vector $ | {w_{x,{\bf{k}}}^ + } \rangle $ is defined as:
\begin{equation}
  \left| {w_{x,{\bf{k}}}^ + } \right\rangle = \sum\limits_{n = 1}^{{N_{occ}}} {\left| {u_{\bf{k}}^n} \right\rangle {{\left[ {\nu _{x,{\bf{k}}}^ + } \right]}^n}} .
\end{equation}
In the thermodynamic limit, The nested Wilson loop we obtained here is just a complex number with module one. We can directly extract observable $p_y^{\nu_x^+}(k_x)$ from the phase factor which describes the polarization in $y$-direction within the Wannier band $\nu_x^+$. And the total Wannier-sector polarization $p_y^{\nu_x^+}$ is obtained by summing over all the $p_y^{\nu_x^+}(k_x)$ with different $k_x$ and divided by $N_x$. With the same procedure, we can derive all the Wannier-sector poloarizations $p_x^{\nu_y^\pm}$, $p_y^{\nu_x^\pm}$ which are able to distinguish different phases in our model.

\section{\label{S2} quantized topological index due to symmetry}

As we have known, in the 1D Su-Schrieffer-Heeger (SSH) model, we can calculate Wannier states and Wannier centers of occupied bands from Wilson loops:
\begin{equation}
  \mathcal{W}_{x,k_x}\!\left| {\nu _{x,{k_x}}^{}} \right\rangle  = \exp \left[ {i2\pi \nu _x} \right]\left| {\nu _{x,{k_x}}^{}} \right\rangle ,  {\nu _x} \in [ - \frac{1}{2}, + \frac{1}{2}].
\end{equation}
The bulk charge polarization ${\bf{p}}_x$ simply equals to the value of the Wannier center $\nu_x$ , which is quantized to $0$ or $1/2$ due to the chiral symmetry. Without symmetry breaking, topological phases with ${\bf{p}}_x = 1/2$ could be distinguished from trivial phases ${\bf{p}}_x=0$, which means you can not adiabatically evolve from one phase to the other without energy band gap closing. Quadrupole insulators studied here can be also understood by a similar way. 

Back to our model, $M_x$, $M_y$ impose constraints on allowed values of four Wannier-sector polarizations defined in \ref{S1}:
\begin{equation}
  \begin{split}
  {M_x}:p_x^{\nu _y^ \pm } &= 0{\rm{ \ or \ }}1/2,
  \\
  {M_y}:p_y^{\nu _x^ \pm } &= 0{\rm{ \ or \ }}1/2.
  \end{split}
\end{equation}
Recall that in 2D systems, inversion $\mathcal{I}$ and $C_2$ transform the coordinates in the same way. Another constraint on the Wannier-sector polarizations due to $\mathcal{I}$ (since $C_2$ symmetry can be obtained by combining two mirror operators $M_x M_y$ in 2D systems) is given as follows:
\begin{equation}
  \mathcal{I}:p_x^{\nu _y^ + } =  - p_x^{\nu _y^ - },p_y^{\nu _x^ + } =  - p_y^{\nu _x^ - }.
\end{equation}
We can learn from the above realtions that our system possesses zero bulk polarizations in both $x$, $y$ directions, which is the premise of defining quadrupole phases. And we can simply use a vector ${{\bf{P}}^\nu } = ( {p_x^{\nu _y^ - },p_y^{\nu _x^ - }} )$ to indicate different topological phases, since we can obtain $p_x^{\nu_y^+},p_y^{\nu_x^+}$ using the relations shown above.
\section{\label{S3} details to determine the critical point and critical exponents}

\begin{figure}
  \includegraphics[scale=0.3]{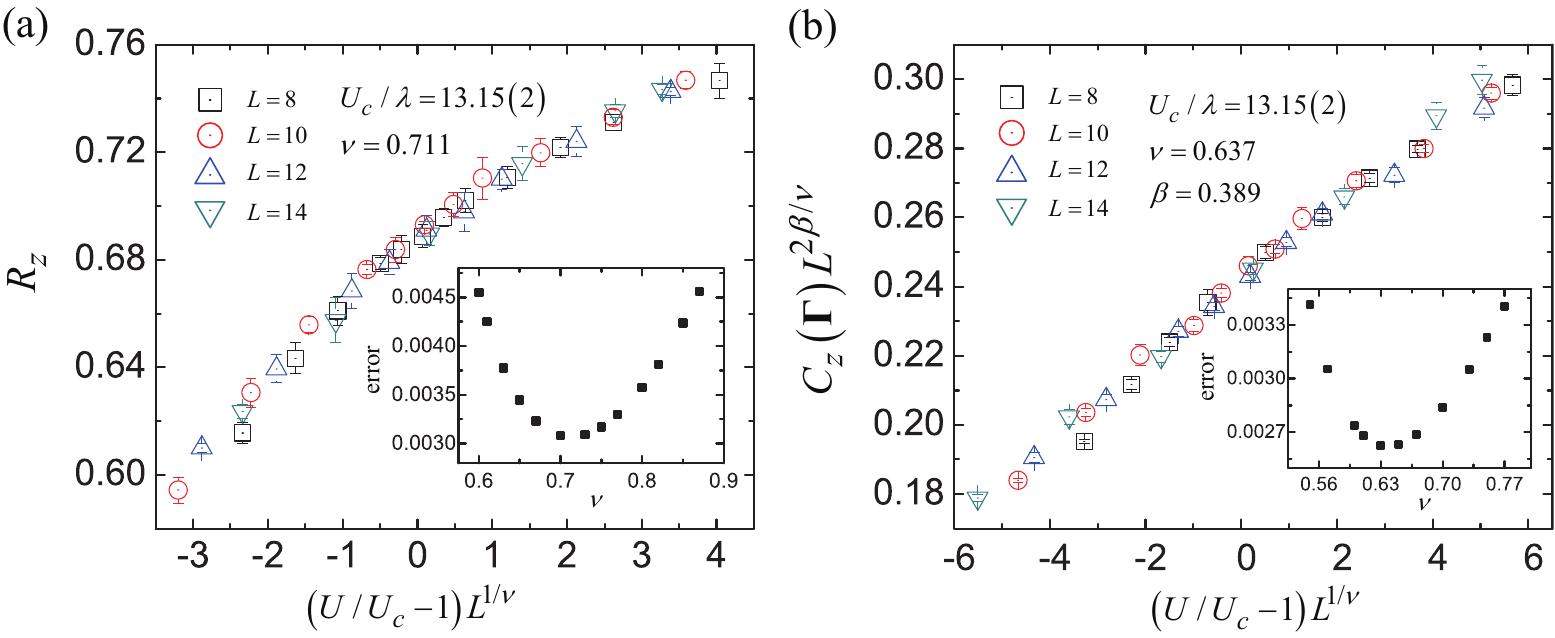} 
  \caption{Data collapse of (a) correlation ratios $R_z(U,L)$ and (b) structure factors $C_z({\bf{\Gamma}})$ with optimized critical exponents. Insets draw the fitting error versus $\nu$ and then give the optimized value of $\nu$ and $\beta$.}
  \label{figs1}
\end{figure}

As we known, it is a tough work to determine the precise value of critical exponents through numerical simulations. In order to derive reliable results, we first make data collapse by assumption $R_z(U,L) = f_R((U/U_c-1)L^{1/\nu})$. Although, the result of regression analysis is insensitive to the variable $\nu$, a precise value of critical point $U_c=13.15(2)$ can be convincingly determined. Then, an optimized value of $\nu=0.711$ can be also obtained by minimizing the error of regression analysis by sweeping $\nu$ (Fig.\ref{figs1}(a)). With a definite value $U_c=13.15$, we can also implement the same idea to find a proper $\beta$ in $C_z(U,L)L^{2\beta/\nu} = f_C((U/U_c-1)L^{1/\nu})$ with different $\nu$. At this time, a pair of value $(\nu=0.637,\beta=0.389)$ can be determined by minimizing the fitting error. Related results are shown in Fig.\ref{figs1}(b) and an error bar of $\nu$ can be given by the deviation between these two minimum points of $\nu$ (inset of Fig.\ref{figs1}(a) and Fig.\ref{figs1}(b)).

\begin{figure}
  \includegraphics[scale=0.33]{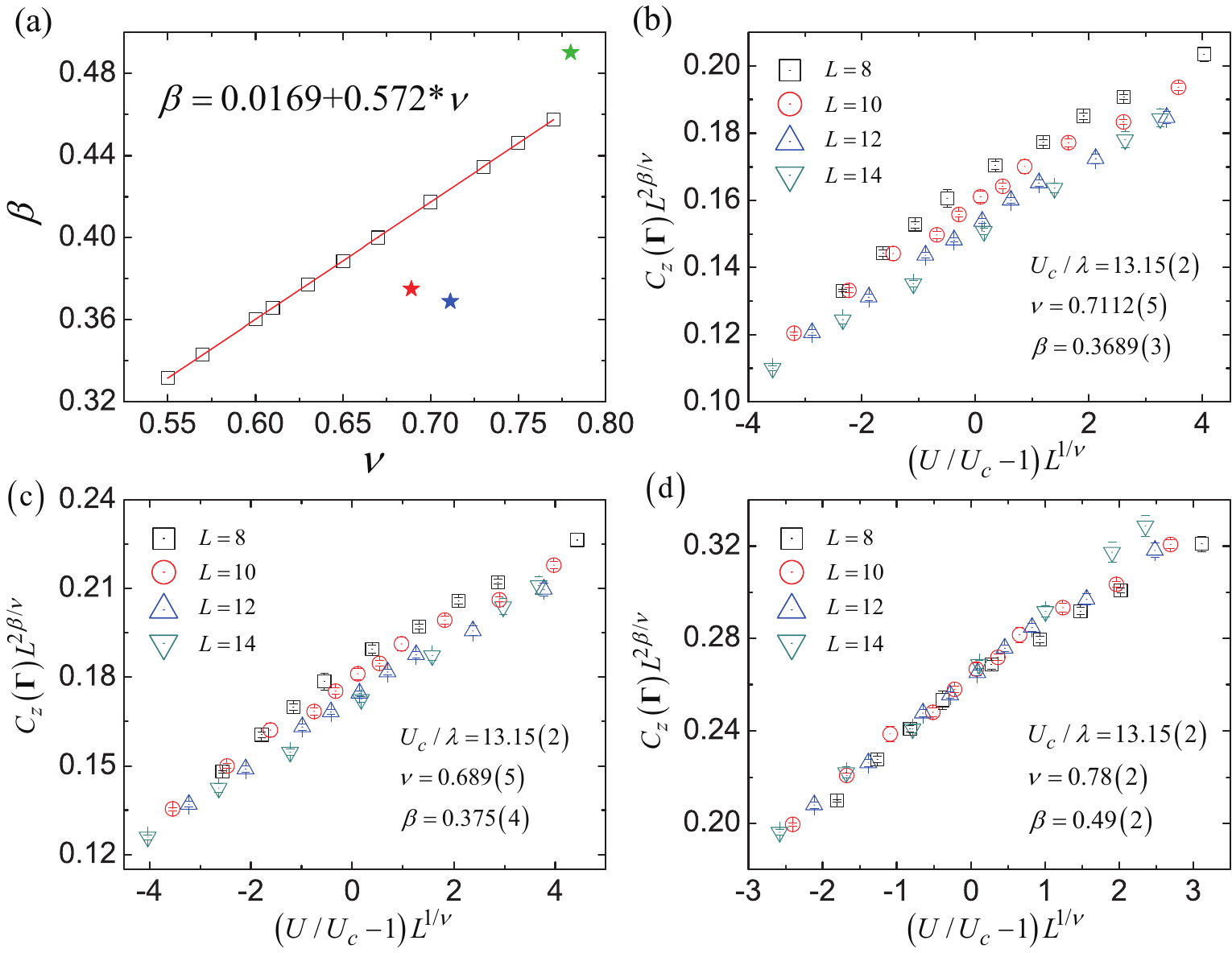} 
  \caption{(a) Black squares stand for the pairs of optimized values $(\nu,\beta)$ around the local minimum of the fitting errors of $C_z(U,L)L^{2\beta/\nu} = f_C((U/U_c-1)L^{1/\nu})$. Critical exponents of 3D Heisenberg universality class, the disorder-AFM transition in dimerized quantum Heisenberg model and the VBS-AFM deconfined quantum criticality are marked by a blue, a red and a green stars in the parameter space, respectively. Data collapses with the critical exponents of (b) 3D Heisenberg universality class \cite{O3class}, (c) the disorder-AFM transition in dimerized quantum Heisenberg model \cite{dimerizedheisenbergmodel} and (d) the VBS-AFM deconfined quantum criticality. \cite{j1j2,jq}.}
  \label{figs2}
\end{figure}

Furthermore, since we can get an optimized $\beta$ by a arbitrary $\nu$ through the regression analysis, we also plot the optimized values of $\beta$ versus $\nu$ in the Fig.\ref{figs2}(a). The relation between $\beta$ and $\nu$ can be described by a linear function $\beta = 0.0169 + 0.572*\nu$. According to our numerical results, we can't give a certain pair of critical exponents $(\nu,\beta)$ since all the points that fall on the line can give a fair fitting error. 
However, we can compared the quantum criticality obtained here with these three cases: (a) 3D Heisenberg universality class, which describes transitions between the AF and a featureless
gapped state, (b) the disorder-AFM transition in dimerized quantum Heisenberg model and (c) the VBS-AFM deconfined quantum criticality found in $J1$-$J2$ and $J$-$Q$ models \cite{O3class,j1j2,dimerizedheisenbergmodel,jq}. The critical exponents of these three universality classes deviate from the line which stands for the optimized exponents of HOTI-AFMI transitions. The distinct deviation is further shown in Fig.\ref{figs2}(b-d) by making data collapses with the exponents corresponding to each of these universality classes, that strongly suggest the TPT investigated in our work is highly non-trivial.

\end{document}